\documentclass[
  ,draft            
  ]
  {aipproc}

\usepackage{xspace}
\layoutstyle{6x9}

\newcommand{\lam}{\ensuremath{\Lambda} \xspace}

\newcommand{\alam}{\ensuremath{\overline{\Lambda}}\xspace}
\newcommand{\kz}{\ensuremath{\mathrm{K^{0}_{S}}}\xspace}

\newcommand{\sqs}{\ensuremath{\sqrt{s}}\xspace}

\newcommand{\pp}{\ensuremath{p+p}\xspace}

\newcommand{\pt}{\ensuremath{p_{T}}\xspace}

\newcommand{\Npart}{\ensuremath{N_{part}}\xspace}

\begin{document}

\title{Strangeness production in small and large collision systems at RHIC}


\author{Mark T.~Heinz (for the STAR Collaboration)}{
  address={Yale University, Wright Nuclear Structure Laboratory,
  272 Whitney Avenue, New Haven, CT 06520} }

\begin{abstract}
We present measurements of strange and multi-strange hadrons in p+p
collisions at $\sqrt{s}$= 200 GeV measured by STAR. We will compare
these preliminary results to leading-order (LO) and next-to-leading
order (NLO) perturbative QCD models widely believed to describe the
production mechanisms. In particular we will point out recent
changes of the model calculations which improve the agreement with
our data significantly and will discuss the physics consequences. In
larger collision systems, produced with heavy ions at RHIC, we
observe the centrality dependence of strange and multi-strange
particle production. The non-linear dependency between
(anti)-hyperon yields and the system size \Npart seems to indicate
that the correlation volume does not scale exactly with \Npart in
contradiction to previous assumptions by thermal models.
\end{abstract}

\maketitle


\section{Introduction}
Perturbative QCD (pQCD) has been successful in describing charged
hadrons production in elementary collisions. Determining the
validity of pQCD in predicting strange particle production in \pp
collisions is important and new high statistics data from the STAR
experiment may provide a useful tool to do so. The 2002 \pp RHIC-run
at $\sqrt{s}$= 200 GeV yielded the largest statistics data on
strange particles at this energy to date. \kz, $\Lambda$, $\Xi$ and
$\Omega$ particles are reconstructed by their secondary vertices in
the STAR Time Projection Chamber (TPC). This paper will compare the
agreement of the transverse momentum spectra observed for \kz,
$\Lambda$ and $\Xi$ with leading-order (\textsc{Pythia}) and NLO
pQCD model calculations \cite{Sjo:87, KKP}.

The data were reconstructed using the large acceptance STAR
experiment at RHIC which is described in more detail elsewhere
\cite{STAR2}. A total of 10.7 million non-singly diffractive (NSD)
events were recorded using the STAR beam-beam counters (BBC) as
triggers. The strange particles were identified from their weak
decay to charged daughter particles. The following decay channels
and the corresponding anti-particles were analyzed: \kz $\rightarrow
\pi^{+} + \pi^{-}$ (b.r. 68.6\%), $\Lambda \rightarrow p +
\pi^{-}$(b.r. 63.9\%), $\Xi^{-} \rightarrow \Lambda + \pi^{-}$(b.r.
99.9\%). Particle identification of the daughters was achieved by
requiring the dE/dx to fall within the 3$\sigma$-bands of
theoretical Bethe-Bloch parameterizations. Further reduction of the
combinatorical background was achieved by applying cuts to select
the decay topology from real decays and full analysis details can be
found elsewhere \cite{MH:thesis}.

\section{Perturbative QCD models}

\begin{figure}[t]
  \includegraphics[height=.23\textheight]{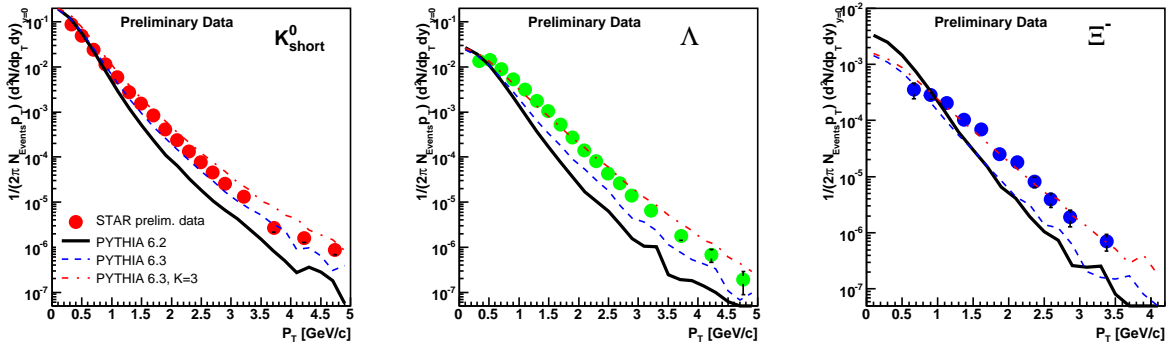}
  \caption{Strange particle transverse momentum spectra compared to 3 different tunes of the PYTHIA pQCD model.
  Errors are statistical only.} \label{fig:Pythia}
\end{figure}

\textsc{Pythia} is a leading order pQCD model based on "Lund" string
fragmentation which describes hadronization. Higher order
corrections to the parton cross-section can be accounted for by
using a phenomenological K-factor. First comparisons to STAR data
were performed using \textsc{Pythia} version 6.220 and results can
be seen in figure 1 and were shown at a previous conference
\cite{MH:HQ04}. In January of 2005, a significantly improved version
6.3 was released by the authors. This version treats multiple
scattering as well as initial and final state showers in a more
consistent way by using a \pt-ordered algorithm \cite{Sjo:04}. The
comparisons of our data to this newer version are also shown in
figure \ref{fig:Pythia}. They clearly show an overall improvement in
the description of the data, however the inclusion of a higher
K-factor, here set to 3, seems necessary to describe the strange and
multi-strange baryons. The K-factor in \textsc{Pythia} simply
multiplies the hard parton cross-section by a constant K. A
discussion of how the K-factor may vary with \sqs can be found
elsewhere \cite{Eskola:03}.

\begin{figure}[ht]
\includegraphics[height=.23\textheight]{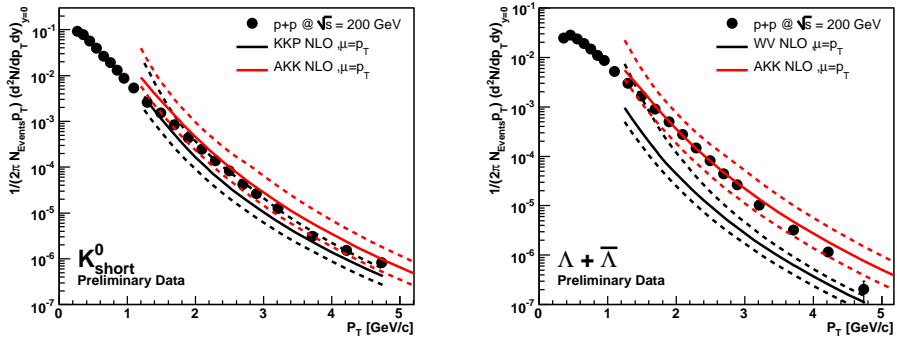}
\caption{Preliminary results for \kz (left) and $\lam+\alam$ (right)
\pt-spectra compared to different NLO calculations. Solid lines are
for $\mu=\pt$, dashed lines are for $\mu=0.5\pt,2\pt$ and indicate
the factorization scale uncertainty. Data includes only statistical
errors.} \label{fig:NLO}
\end{figure}

Next-to-leading order calculations are similar to \textsc{Pythia} in
the sense that they also apply the factorization theorem to separate
the perturbative from the non-perturbative part of the process. The
perturbative parton cross-sections are analytically calculated using
the additional higher order Feynman diagrams that were absent in LO.
The non-perturbative part consisting of the proton parton
distribution function (PDF) and the fragmentation function (FF) are
parametrized using collider data.

Previously, it has been shown that the charged particle and neutral
pion spectra at RHIC are well described by NLO pQCD calculations
\cite{MVL2005}. The fact that the strange baryons require a higher
K-factor in the LO calculation seems to indicate that a more precise
calculation of the parton cross-section is necessary. Figure
\ref{fig:NLO} compares recent NLO calculations to strange particle
spectra measured by STAR. For \kz, two calculations were obtained,
one using Kniehl-Kramer-P\"{o}tter (KKP) fragmentation functions
\cite{KKP}, the other using more recent parametrization by Albino
\emph{et al.} (AKK) \cite{AKK}. Both describe the data within the
model uncertainties. For \lam, calculations by Vogelsang \emph{et
al.} (WV) \cite{WV} and by AKK were used. The more recent AKK
parametrization includes an additional assumption that the shape of
the gluon fragmentation function for \lam is similar to that of the
proton but suppressed by a factor of 3. This factor was derived by
using the STAR data as an additional constraint.

\section{Strangeness in large systems}
Heavy ion collisions at RHIC produce a system of deconfined quarks
and gluons of much larger volume than \pp collisions. The
enhancement of strange particles has been proposed as a signature of
such a strongly interacting Quark-Gluon Plasma. The dependency of
the strangeness yields with respect to the created volume of the
system has been reported in heavy ion collisions at CERN SPS
(\sqs=17 GeV) and studied by statistical models \cite{Redlich:02}.
Recently STAR measured the yields for \lam and $\Xi$ particles in
Au+Au collisions at \sqs=200 GeV for various collision centralities
\Npart \cite{Caines}. The observed non-linear increase of the yields
with \Npart contradicts the SPS measurement and indicates that the
strangeness correlation volume may not be well described by \Npart.
Alternative results by a non-equilibrium statistical model seem to
agree better with the STAR data \cite{Rafelski:05}.

\bibliographystyle{aipproc}   


\end{document}